# SYMMETRIES IN PHYSICS: PHILOSOPHICAL REFLECTIONS

*EDITED BY*
K. BRADING[1] AND E. CASTELLANI[2]

*CAMBRIDGE UNIVERSITY PRESS*, 2003

## CONTENTS

[1] Wolfson College, Oxford. E-mail: katherine.brading@wolfson.ox.ac.uk
[2] University of Florence, Italy. E-mail: castella@unifi.it

# Introduction

KATHERINE BRADING AND ELENA CASTELLANI

This book is about the various symmetries at the heart of modern physics. How should we understand them and the different roles that they play? Before embarking on this investigation, a few words of introduction may be helpful. We begin with a brief description of the historical roots and emergence of the concept of symmetry that is at work in modern physics (section 1). Then, in section 2, we mention the different varieties of symmetry that fall under this general umbrella, outlining the ways in which they were introduced into physics. We also distinguish between two different uses of symmetry: symmetry principles versus symmetry arguments. In section 3 we change tack, stepping back from the details of the various symmetries to make some remarks of a general nature concerning the status and significance of symmetries in physics. Finally, in section 4, we outline the structure of the book and the contents of each part.

## 1 *The meanings of symmetry*

Symmetry is an ancient concept. Its history starts with the Greeks, the term summetria deriving from sun (with, together) and metron (measure) and originally indicating a relation of commensurability (such is the meaning codified in Euclid's *Elements*, for example). But symmetry immediately acquired a further, more general meaning, with commensurability representing a particular case: that of a proportion relation, grounded on (integer) numbers, and with the function of harmonizing the *different* elements into a *unitary whole*: "The most beautiful of all links is that which makes, of itself and of the things it connects, the greatest unity possible; and it is the proportion (summetria) which realizes it in the most beautiful way" (Plato, *Timaeus*, 31c).

From the outset, then, symmetry was closely related to harmony, beauty, and unity, and this was to prove decisive for its role in theories of nature. In Plato's *Timaeus*, for example, the regular polyhedra are afforded a central place in the doctrine of natural elements for the proportions they contain and the beauty of their forms: fire has the form of the regular tetrahedron, earth the form of the cube, air the form of the regular octahedron, water the form of the regular icosahedron, while the regular dodecahedron is used for the form of the entire universe. The history of science provides another paradigmatic example of the use of these figures as basic ingredients in physical description: Kepler's 1596 *Mysterium cosmographicum* presents a planetary architecture grounded on the five regular solids.

The regular figures used in Plato's and Kepler's physics for the mathematical proportions and harmonies they contain (and the related properties and beauty of their form) are symmetric in another sense that does not have to do with proportions. In the language of modern science, the symmetry of geometrical figures – such as the regular polygons and polyhedra – is defined in terms of their invariance under specified groups of rotations and reflections. Where does this definition stem from? Besides the ancient notion of symmetry used by the Greeks and Romans (current until the end of the Renaissance), a different notion of symmetry slowly emerged in the modern era, grounded not on proportions but on an equality relation. More precisely, it is grounded on an equality relation between elements that are opposed, such as the left and right parts of a figure. This notion, explicitly recognized and defined in such terms in a 1673 text by Claude Perrault, is, in fact, nothing other than our reflection symmetry. Reflection symmetry has now a precise definition in terms of invariance under the group of reflections, representing a particular case of the group-theoretic notion of symmetry currently used in modern science.

In moving from Perrault's notion to this abstract group-theoretic notion the following crucial steps are worth noting. First, we have the interpretation of the *equality of the parts with respect to the whole* in the sense of their *interchangeability* (equal parts can be exchanged with one another, while preserving the whole). Then, we have the introduction of specific mathematical operations, such as reflections, rotations, and translations, that are used to describe with precision how the parts are to be exchanged. As a result, we arrive at a definition of the symmetry of a geometrical figure in terms of its *invariance* when equal component parts are exchanged according to one of the specified operations. Thus, when the two halves of a bilaterally symmetric figure are exchanged by reflection, we recover the original figure, and that figure is said to be invariant under left–right reflections. This is known as the "crystallographic notion of symmetry", since it was in the context of early developments in crystallography that symmetry was first so defined and applied. The next key step is the generalization of this notion to the *group-theoretic* definition of symmetry, which arose following the nineteenth-century development of the algebraic concept of a group, and the fact that the symmetry operations of a figure were found to satisfy the conditions for forming a group.[3] Finally, as is discussed in more detail later in this volume (see Castellani, section IV), we have the resulting close connection between the notion of symmetry, equivalence and group (a symmetry group induces a partition into equivalence classes).

---

[3] A group is defined to be a set G, together with a product operation (•), such that: for any two elements $g_1$ and $g_2$ of G, $g_1 • g_2$ is again an element of G; the group operation is associative; the group contains the identity element; and for each element there exists an inverse.

The group-theoretic notion of symmetry is the one that has proved so successful in modern science, and with which the papers of this collection are concerned. Note, however, that symmetry remains linked to beauty (regularity) and unity: by means of the symmetry transformations, distinct (but "equal" or, more generally, "equivalent") elements are related to each other and to the whole, thus forming a regular "unity". The way in which the regularity of the whole emerges is dictated by the nature of the specified transformation group. Summing up, a *unity of different and equal elements* is always associated with symmetry, in its ancient or modern sense; the way in which this unity is realized, on the one hand, and how the equal and different elements are chosen, on the other hand, determines the resulting symmetry and in what exactly it consists.[4]

## 2  *Symmetry in the history of physics*

When considering the role of symmetry in physics from a historical point of view, it is worth keeping in mind two preliminary distinctions:

➢ The first is between implicit and explicit uses of the notion. Symmetry considerations have always been applied to the description of nature, but for a long time in an implicit way only. As we have seen, the scientific notion of symmetry (the one we are interested in here) is a recent one. If we speak about a role of this concept of symmetry in the ancient theories of nature, we must be clear that it was not used explicitly in this sense at that time.

➢ The second is between the two main ways of using symmetry. First, we may attribute specific symmetry properties to physical situations or phenomena, or to laws (*symmetry principles*). It is the application with respect to *laws*, rather than to objects or phenomena, that has become central to modern physics, as we will see. Second, we may derive specific consequences with regard to particular physical situations or phenomena on the basis of their symmetry properties (*symmetry arguments*).

**Symmetry principles**

Nature offers plenty of examples of (approximate) symmetrical forms: the bilateral symmetry of the human (and, in general, of animal) bodies, the pentagonal symmetry frequent in flowers, the hexagonal symmetry of honeycomb cells, the translational symmetry of plant shoots and of animals such as caterpillars, and so on. The natural objects with the richest and most evident symmetry properties are undoubtedly crystals, and so it is not surprising that the systematic study of all possible symmetric configurations – the so-called theory of symmetry – started in connection with the rise of crystallography. The classification of all symmetry properties of crystals, which produced its most notable results in the nineteenth century, in fact marks the first explicit application of the scientific notion of symmetry in science.[5] The real turning point in the use of symmetry in science came, however, with the introduction of the *group* concept and with the ensuing developments in the theory of transformation groups. This is because the group-theoretic definition of symmetry as "invariance under a specified group of transformations" allowed the concept to be applied much more widely, not only to spatial figures but also to abstract objects such as mathematical expressions – in particular, expressions of physical relevance such as dynamical equations. Moreover, the technical apparatus of group theory could then be transferred and used to great advantage within physical theories.

---

[4] Further details of the material in this section can be found in Castellani (2000), chapters 1-3.

[5] Symmetry considerations were used by Haüy to characterize and classify crystal structure and formation (see his 1801 *Traité de minéralogie*, volume 1), and with this, crystallography emerged as a discipline distinct from mineralogy. From Haüy's work two strands of development led to the 32 point transformation crystal classes and the 14 Bravais lattices, all of which may be defined in terms of discrete groups. These were combined into the 230 space groups by Fedorov and by Schönflies in 1891, and by Barlow in 1894. The theory of discrete groups continues to be fundamental in solid state physics, chemistry, and material science.

The first explicit study of the invariance properties of equations in physics is connected with the introduction, in the first half of the nineteenth century, of the transformational approach to the problem of motion in the framework of analytical mechanics. Using the formulation of the dynamical equations of mechanics due to Hamilton (known as the Hamiltonian or canonical formulation), Jacobi developed a procedure for arriving at the solution of the equations of motion based on the strategy of applying transformations of the variables that leave the Hamiltonian equations invariant, thereby transforming step by step the original problem into new ones that are simpler but perfectly equivalent (for further details see Lanczos, 1949).[6] Jacobi's canonical transformation theory, although introduced for the "merely instrumental" purpose of solving dynamical problems, led to a very important line of research: the general study of physical theories in terms of their transformation properties. Examples of this are the studies of invariants under canonical transformations, such as Poisson brackets or Poincaré's integral invariants; the theory of continuous canonical transformations due to Lie; and, finally, the connection between the study of physical invariants and the algebraic and geometric theory of invariants that flourished in the second half of the nineteenth century, and which laid the foundation for the geometrical approach to dynamical problems. The use of the mathematics of group theory to study physical theories was central to the work, early in the twentieth century in Göttingen, of the group whose central figures were Klein (who earlier collaborated with Lie) and Hilbert, and which included Weyl and later Noether. We will return to Weyl and Noether later.

On the above approach, the equations or expressions of physical interest are already given, and the strategy is to study their symmetry properties. There is, however, an alternative way of proceeding, namely the reverse one: start with specific symmetries and search for dynamical equations with such properties. In other words, we *postulate* that certain symmetries are physically significant, rather than deriving them from prior dynamical equations. The assumption of certain symmetries in nature is not, of course, a novelty. Although not explicitly expressed as symmetry principles, the homogeneity and isotropy of physical space, and the uniformity of time (forming together with the invariance under Galilean boosts "the older principles of invariance" – see Wigner, 1967; also in this volume, section IV), have been assumed as prerequisites in the physical description of the world since the beginning of modern science. Perhaps the most famous early example of the deliberate use of this type of symmetry principle is Galileo's discussion of whether the Earth moves, in his *Dialogue concerning the two chief world systems* of 1632. Galileo sought to neutralize the standard arguments purporting to show that, simply by looking around us at how things behave locally on Earth – how stones fall, how birds fly – we can conclude that the Earth is at rest rather than rotating, arguing instead that these observations do not enable us to determine the state of motion of the Earth. His approach was to use an analogy with a ship: he urges us to consider the behaviour of objects, both animate and inanimate, inside the cabin of a ship, and claims that no experiments carried out inside the cabin, without reference to anything outside the ship, would enable us to tell whether the ship is at rest or moving smoothly across the surface of the Earth. The *assumption* of a symmetry between rest and a certain kind of motion leads to the prediction of this result, without the need to know the laws governing the experiments on the ship. The "Galilean principle of relativity" (according to which the laws of physics are invariant under Galilean boosts, where the states of motion considered are now those of uniform velocity) was quickly adopted as an axiom and widely used in the seventeenth century, notably by Huygens in his solution to the problem of colliding bodies and by Newton in his early work on motion. Huygens took the relativity principle as his 3rd hypothesis or axiom, but in Newton's *Principia* it is demoted to a corollary to the laws of motion, its status in Newtonian physics therefore being that of a *consequence* of the laws, even though it remains, in fact, an independent assumption.

---

[6] Notice that this is a clear example of a methodological use of symmetry properties: on the basis of the invariance properties of the situation under consideration (in this case, the dynamical problem in classical mechanics), a strategy is applied for deriving determinate consequences. The underlying principle is that equivalent problems have equivalent solutions. This type of symmetry argument (see section 0, below) is discussed also by van Fraassen (1989), chapter 10.

Although the spatial and temporal invariance of mechanical laws were known and used for a long time in physics, and the group of the global spacetime symmetries for electrodynamics was completely derived by Poincaré[7] before Einstein's famous 1905 paper setting out his special theory of relativity, it was not until this work by Einstein that the status of symmetries with respect to the laws was reversed. Wigner (1967; see this volume, section IV) writes that "the significance and general validity of these principles were recognized, however, only by Einstein", and that Einstein's work on special relativity marks "the reversal of a trend: until then, the principles of invariance were derived from the laws of motion … It is now natural for us to derive the laws of nature and to test their validity by means of the laws of invariance, rather than to derive the laws of invariance from what we believe to be the laws of nature". In postulating the universality of the global continuous spacetime symmetries – also known as "geometrical symmetries" in the terminology introduced by Wigner (*ibid.*; see this volume, section I) – Einstein's construction of his special theory of relativity represents the first turning point in the application of symmetry to twentieth-century physics.[8]

Global spacetime invariance principles are intended to be valid for all the laws of nature. Such a universal character is not shared by the physical symmetries that were next introduced in physics. Most of these were of an entirely new kind, with no roots in the history of science, and in some cases expressly introduced to describe specific forms of interactions – whence the name "dynamical symmetries" due to Wigner (*ibid.*; see this volume, section I).

The new symmetries were for the most part closely related to specific features of the microscopic world. *Permutation symmetry*, "discovered" by Heisenberg in 1926 in relation to the indistinguishability of so-called identical quantum particles (see French and Rickles, this volume), was the first non-spatiotemporal symmetry to be introduced into microphysics, and also the first symmetry to be treated with the techniques of group theory in the context of quantum mechanics. The application of the theory of groups and their representations for the exploitation of symmetries in quantum mechanics undoubtedly represents the second turning point in the twentieth-century history of physical symmetries. It is, in fact, in the quantum context that symmetry principles are at their most effective. Wigner and Weyl were among the first to recognize the great relevance of symmetry groups to quantum physics and the first to reflect on the meaning of this. As Wigner emphasized on many occasions, one essential reason for the "increased effectiveness of invariance principles in quantum theory" (Wigner, 1967, p. 47) is the linear nature of the state space of a quantum physical system, corresponding to the possibility of superposing quantum states. This gives rise to, among other things, the possibility of defining states with particularly simple transformation properties in the presence of symmetries.

---

[7] Whence the name 'Poincaré group' introduced later by Wigner, whereas Poincaré himself named the group after Lorentz.

[8] General relativity marks a further important stage in the development, as we will see below. See also Martin, this volume.

In general, if G is a symmetry group of a theory describing a physical system (that is, the fundamental equations of the theory are invariant under the transformations of G), this means that the states of the system transform into each other according to some "representation" of the group G. In other words, the group transformations are mathematically represented in the state space by operations relating the states to each other. In quantum mechanics, these operations are generally the operators acting on the state space that correspond to the physical observables, and any state of a physical system can be described as a superposition of states of elementary systems, that is, of systems the states of which transform according to the "irreducible" representations of the symmetry group. Quantum mechanics thus offers a particularly favourable framework for the application of symmetry principles. The observables representing the action of the symmetries of the theory in the state space, and therefore commuting with the Hamiltonian of the system, play the role of the conserved quantities; furthermore, the basis states may be labelled by the irreducible representations of the symmetry group, which accordingly also regulate the transformations from one state to another (state transitions).

But more can be said. Because of the specific properties of the quantum description, symmetries such as spatial reflection symmetry or *parity* (P**)** and *time reversal* (T) were "rediscovered" in the quantum context, taking on a new significance.[9] Moreover, new "quantum symmetries" emerged, such as particle–antiparticle symmetry or *charge conjugation* (C),[10] and the various *internal symmetries* grounded on invariances under phase changes of the quantum states and described in terms of the unitary groups SU(N) (the local versions of which are the *gauge symmetries* at the core of the Standard Model for elementary particles).[11] More recently, new symmetries acquired relevance in theoretical physics, such as *supersymmetry* (the symmetry relating bosons and fermions and leading, when made local, to the theories of *supergravity*), and the various forms of *duality* used in today's superstring theories.

The history of the application of symmetry principles in quantum mechanics and then quantum field theories coincides with the history of the developments of twentieth-century theoretical physics. The salient aspects of this history, from the perspective of the meaning of physical symmetries, are discussed in the contributions to this volume (for details, see section 0, below) and cover four crucial developments.

---

[9] Parity was introduced in quantum physics in 1927 in a paper by Wigner, where important spectroscopic results were explained for the first time on the basis of a group-theoretic treatment of permutation, rotation and reflection symmetries. Time reversal invariance appeared in the quantum context, again due to Wigner, in a 1932 paper .

[10] Charge conjugation was introduced in Dirac's famous 1931 paper 'Quantized singularities in the electromagnetic field'. C is a discrete symmetry, connected to the spatial and temporal discrete symmetries P and T by the so-called CPT theorem, demonstrated by Lüders in 1952, which states that the combination of C, P, and T is a general symmetry of physical laws.

[11] The starting point for the idea of internal symmetries was the interpretation of the presence of particles with (approximately) the same value of mass as the components (*states*) of a single physical system, connected to each other by the transformations of an underlying symmetry group. This idea emerged in analogy with what happened in the case of permutation symmetry, and was in fact due to Heisenberg (the discoverer of permutation symmetry), who in a 1932 paper introduced the SU(2) symmetry connecting the proton and the neutron (interpreted as the two states of a single system). This symmetry was further studied by Wigner, who in 1937 introduced the term *isotopic spin* (later contracted to *isospin*).

- ➢ The first is the extension of the concept of continuous symmetry from "global" symmetries (such as the Galilean group of spacetime transformations) to "local" symmetries, as discussed by Martin (this volume) in his review of continuous symmetries. Einstein was the first to make use of a local symmetry principle in theory construction when developing his General Theory of Relativity (GTR), culminating in 1915.[12] Meanwhile in Göttingen, Klein and Hilbert enlisted the assistance of Noether in their investigations into the status of energy conservation in generally covariant theories of gravitation. This led to Noether's famous 1918 paper containing two theorems, the first of which leads to a connection between global symmetries and conservation laws, and the second of which allows a demonstration of the different status of these conservation laws when the global symmetry group is a subgroup of some local symmetry group of the theory in question (see Brading and Brown, this volume). Prompted by Einstein's work, Weyl's 1918 "unified theory of gravitation and electromagnetism" extended the idea of local symmetries (see Ryckman, this volume), and although this theory is generally deemed to have failed, the theory contains the seeds of later success in the context of quantum theory (see below).
- ➢ The second is the extension of the concept of continuous symmetry from spatiotemporal to internal, both global and local. In quantum theory, the phase of the wavefunction encodes internal degrees of freedom. With the requirement that a theory be invariant under *local gauge transformations* involving the phase of the wavefunction, Weyl's ideas found a successful home in quantum theory (see O'Raifeartaigh, 1997). Weyl's new 1929 theory was a theory of electromagnetism coupled to matter. The history of gauge theory is surveyed briefly by Martin (this volume), who highlights various issues surrounding gauge symmetry, in particular the status of the so-called "gauge principle", first proposed by Weyl. Martin also discusses the ensuing stages in the development of gauge theory, the main steps being the Yang and Mills non-Abelian gauge theory of 1954, and the problems and solutions associated with the successful development of gauge theories for the short-range weak and strong interactions.
- ➢ The third is the increasing importance of the discrete symmetries of permutation invariance and C, P, and T mentioned above.
- ➢ Finally, the fourth is the introduction in the late 1950s and early 60s of the concept of spontaneous symmetry breaking in field theory (see section III of this volume), and the subsequent related results (including the Goldstone 1961 theorem and the 1964 so-called Higgs mechanism), which played a crucial role in the developments of the Standard Model of elementary particles.

**Symmetry arguments**

Consider the following cases.

- ➢ Buridan's ass: situated between what are, for him, two completely equivalent bundles of hay, he has no reason to choose the one located to his left over the one located to his right, and so he is not able to choose and dies of starvation.

---

[12] See Norton (this volume) on the "Kretschmann objection" to the physical significance of general covariance, and also Martin (this volume), section 2.2, on invariance versus covariance.

- Archimedes's equilibrium law for the balance: if equal weights are hung at equal distances along the arms of a balance, then it will remain in equilibrium since there is no reason for it to rotate one way or the other about the balance point.
- Anaximander's argument for the immobility of the Earth as reported by Aristotle: the Earth remains at rest since, being at the centre of the spherical cosmos (and in the same relation to the boundary of the cosmos in every direction), there is no reason why it should move in one direction rather than another.

What do they have in common?

First, these can all be understood as examples of the application of the Leibnizean Principle of Sufficient Reason (PSR): if there is no sufficient reason for one thing to happen instead of another, the principle says that nothing happens (the initial situation does not change). But there is something more that the above cases have in common: in each of them PSR is applied on the grounds that the initial situation has a given symmetry: in the first two cases, bilateral symmetry; in the third, rotational symmetry. The symmetry of the initial situation implies the complete equivalence between the existing alternatives (the left bundle of hay with respect to the right one, and so on). If the alternatives are completely equivalent, then there is no sufficient reason for choosing between them and the initial situation remains unchanged.

Arguments of the above kind – that is, arguments leading to definite conclusions on the basis of an initial symmetry of the situation plus PSR – have been used in science since antiquity (as Anaximander's argument testifies). The form they most frequently take is the following: a situation with a certain symmetry evolves in such a way that, in the absence of an asymmetric cause, the initial symmetry is preserved. In other words, a breaking of the initial symmetry cannot happen without a reason, or *an asymmetry cannot originate spontaneously*. Van Fraassen (1989) devotes a chapter to considering the way these kinds of symmetry arguments can be used in general problem-solving.

Historically, the first explicit formulation of this kind of argument in terms of symmetry is due to the physicist Pierre Curie towards the end of nineteenth century. Curie was led to reflect on the question of the relationship between *physical properties* and *symmetry properties* of a physical system by his studies on the thermal, electric and magnetic properties of crystals, these properties being directly related to the structure, and hence the symmetry, of the crystals studied. More precisely, the question he addressed was the following: in a given physical medium (for example, a crystalline medium) having specified symmetry properties, which physical phenomena (for example, which electric and magnetic phenomena) are allowed to happen? His conclusions, systematically presented in his 1894 work "Sur la symétrie dans les phénomènes physiques" (see this volume, section III), can be synthesized as follows:

(a) A phenomenon can exist in a medium possessing its characteristic symmetry or that of one of its subgroups. What is needed for its occurrence (i.e. for something rather than nothing to happen) is not the presence, but rather the absence, of certain symmetries: "Asymmetry is what creates a phenomenon".
(b) The symmetry elements of the causes must be found in their effects, but the converse is not true; that is, the effects can be more symmetric than the causes.

Conclusion (a) clearly indicates that Curie recognized the important function played by the concept of symmetry breaking in physics (he was indeed one of the first to recognize it). Conclusion (b) is what is usually called "Curie's principle" in the literature, although notice that (a) and (b) are not independent of one another.

In order for Curie's principle to be applicable, various conditions need to be satisfied: the causal connection must be valid, the cause and effect must be well-defined, and the symmetries of both the cause and the effect must also be well-defined (this involves both the physical and the geometrical properties of the physical systems considered). Curie's principle then furnishes a *necessary condition* for given phenomena to happen: only those phenomena can happen that are compatible with the symmetry conditions established by the principle.

Curie's principle has thus an important methodological function: on the one side, it furnishes a kind of selection rule (given an initial situation with a specified symmetry, only certain phenomena are allowed to happen); on the other side, it offers a falsification criterion for physical theories (a violation of Curie's principle may indicate that something is wrong in the physical description).[13]

Such applications of Curie's principle depend, of course, on our accepting its validity, and this is something that has been questioned in the literature, especially in relation to spontaneous symmetry breaking (see this volume, section III). Different proposals have been offered for justifying the principle. We have presented it here as an example of symmetry considerations based on Leibniz's PSR, while Curie himself seems to have regarded it as a form of causality principle. Chalmers (1970) considers its relation to the invariance properties of physical laws and argues that the principle follows from these in the case of deterministic laws, a point of view taken up again and generalized in Ismael (1997). On this approach, Curie's principle is understood as a condition on the relationship between the symmetries of a problem (an equation) and its solution(s). This has the advantages of avoiding the apparent vagueness of Curie's formulation (the appeal to causality, and so forth) while also extending it to cover symmetries of physical laws. However, trying to generalize Curie's principle as a principle about the link between the symmetries of an equation and its solution(s) is not straightforward and requires further attention (for more on symmetries of laws versus symmetries of solutions, see Belot, this volume, and Castellani, this volume, section III).

### 3  *Symmetries of modern physics: their status and significance*

What is the status and significance of symmetries and symmetry principles in physics? The rich variety of symmetries in modern physics means that such a general question is not easily addressed. Indeed, we might even wonder whether it is well posed, and restrict our questions instead to specific symmetries and the interpretational issues they raise. Much of the recent literature opts for such restrictions on scope, and this is reflected in sections I–III of this book (see also section 0 of this introduction). However, something can be said in more general terms; here we offer a few remarks in that direction[14] and we refer the reader to section IV of the book, where general interpretative issues are addressed.

Exploring the roles and meanings of symmetries is deeply intertwined with basic questions regarding physical reality and physical knowledge, along with the methodologies and guiding strategies of contemporary physical inquiry. Thus, in approaching the above question we must take into account the possible ontological, epistemological, and methodological aspects of symmetries. In order to do this, we think that it is helpful to begin by considering the different roles that symmetries play in physics, the main four being, in our opinion, classificatory, normative, unifying, and explanatory.

---

[13] See, for example, Mach's discussion of the Oersted effect in his *Die Mechanik in ihrer Entwickelung historisch – kritisch dargestellt* of 1883.

[14] This section of the introduction is based on Castellani (2002).

One of the most important roles played by symmetry is that of *classification* – for example, the classification of crystals using their remarkable and varied symmetry properties. In contemporary physics, the best example of this role of symmetry is the classification of elementary particles by means of the irreducible representations of the fundamental physical symmetry groups, a result first obtained by Wigner in his famous paper of 1939 on the unitary representations of the inhomogeneous Lorentz group. If a symmetry classification includes all the necessary properties for characterizing a given type of physical object (for example, all necessary quantum numbers for characterizing a given type of particle), we have the possibility of defining types of entities on the basis of their transformation properties. This has led philosophers of science to explore a structuralist approach to the entities of modern physics, in particular a group-theoretical account of objects (see for example the contributions in Castellani, 1998, part II).

Symmetries also have a *normative* role, being used as constraints on physical theories. The requirement of invariance with respect to a transformation group imposes severe restrictions on the form that a theory may take, limiting the types of quantities that may appear in the theory as well as the form of its fundamental equations. A famous case is Einstein's use of general covariance when searching for his gravitational equations.

The group-theoretical treatment of physical symmetries, with the resulting possibility of unifying different types of symmetries by means of a unification of the corresponding transformation groups, has provided the technical resources for symmetry to play a powerful role in theoretical *unification*. This is best illustrated by the current – dominant – research programme in theoretical physics aimed at arriving at a unified description of all the fundamental forces of nature (gravitational, weak, electromagnetic and strong) in terms of underlying local symmetry groups.

It is often said that many physical phenomena can be explained as (more or less direct) consequences of symmetry principles or symmetry arguments. In the case of symmetry principles, the *explanatory* role of symmetries arises from their place in the hierarchy of the structure of physical theory, which in turn derives from their generality. For example, an explanatory role for symmetries with respect to conservation laws might be claimed on the basis of Noether's connection between symmetries and conservation laws (see Brading and Brown, this volume), along with the more fundamental status of symmetries in the hierarchy. As Wigner describes the hierarchy (Wigner, 1967; see especially the second extract in section IV of this volume), symmetries are seen as properties of the laws. Thus, through the requirement that the laws (whatever they may be) must be invariant under certain symmetries, these symmetries place constraints on which events are physically possible (the explanatory role clearly connects to the normative role here). In other words, symmetries may be used to explain (i) the form of the laws, and (ii) the occurrence (or non-occurrence) of certain events (this latter in a manner analogous to the way in which the laws explain why certain events occur and not others). Other features of symmetry in physics that are commonly used as an important explanatory basis for physical phenomena are the "gauge principle" (for the form, or even existence, of the various interactions; see Martin, this volume) and the mechanism of "spontaneous symmetry breaking" (see this volume, section III). Finally, insofar as explanatory power may be derived from unification, the unifying role of symmetries also results in an explanatory role.

In the latter case, i.e. that of symmetry arguments, we may, for example, appeal to Curie's principle to explain the occurrence of certain phenomena on the basis of the symmetries (or asymmetries) of the situation, as discussed in section 2.2, above.

From these different roles we can draw some preliminary conclusions about the status of symmetries. It is immediately apparent that symmetries have an important *heuristic* function, indicating a strong *methodological* status. What about the ontological and epistemological status of symmetries?

Adopting an *ontological* view, symmetries are seen as a substantial part of the physical world: the symmetries of theories represent properties existing in nature, or characterize the structure of the physical world. It might be claimed, furthermore, that the ontological status of symmetries provides the reason for the methodological success of symmetries in physics. A concrete example is the use of symmetries to predict the existence of new particles. This can happen via the *classificatory* role, on the grounds of vacant places in symmetry classification schemes, as in the famous case of the 1962 prediction of the particle     in the context of the hadronic classification scheme known as the "Eightfold Way". Or, as in more recent cases, via the *unificatory* role: the paradigmatic example is the prediction of the W and Z particles (experimentally found in 1983) in the context of the Weinberg-Salam gauge theory proposed in 1967 for the unification of the weak and electromagnetic interactions.[15] These impressive cases of the prediction of new phenomena might perhaps be used to argue for an *ontological* status for symmetries, via an inference to the best explanation.

Another reason for attributing symmetries to nature is the so-called geometrical interpretation of spatiotemporal symmetries, according to which the spatiotemporal symmetries of physical laws are interpreted as symmetries of spacetime itself, the "geometrical structure" of the physical world. Moreover, this way of seeing symmetries can be extended to non-external symmetries, by considering them as properties of other kinds of spaces, usually known as "internal spaces". The question of exactly what a realist would be committed to on such a view of internal spaces remains open, and an interesting topic for discussion – in this regard see Nounou, this volume.

One approach to investigating the limits of an ontological stance with respect to symmetries would be to investigate their empirical or observational status: can the symmetries in question be directly observed? Morrison (this volume) raises concerns about a realist approach for the case of spontaneously broken symmetries, and the question can also be tackled for symmetries that are not spontaneously broken. We first have to address what it means for a symmetry to be observable, and indeed whether all (unbroken) symmetries have the same observational status. Kosso (2000) arrives at the conclusion that there are important differences in the empirical status of the different kinds of symmetries. In particular, while global continuous symmetries can be directly observed – via such experiments as the Galilean ship experiment – a local continuous symmetry can have only indirect empirical evidence. Brading and Brown[16] argue for a different interpretation of Kosso's examples,[17] and hence for a different understanding of why the local symmetries of gauge theory and GTR have an empirical status distinct from that of the familiar global spacetime symmetries. The most fundamental point is this: in theories with local gauge symmetry, the matter fields are embedded in a gauge field, and the local symmetry is a property of both sets of fields *jointly*. Because of this there is, in general, no analogue of the Galilean ship experiment for local symmetry transformations; according to Brading and Brown, the continuous global spacetime symmetries have a special empirical status.[18]

---

[15] The unificatory role of symmetries in physics is associated with a more general realist metaphysics influential amongst theoretical physicists working towards a unified theory of everything.
[16] K. A. Brading and H. R. Brown, "Observing gauge symmetry transformations?", ms.
[17] Kosso's analysis begins from a set of examples offered by 't Hooft (1980).
[18] See also Brading and Brown, this volume.

The direct observational status of the familiar global spacetime symmetries leads us to an *epistemological* aspect of symmetries. According to Wigner, the spatiotemporal invariance principles play the role of a prerequisite for the very possibility of discovering the laws of nature: "if the correlations between events changed from day to day, and would be different for different points of space, it would be impossible to discover them" (Wigner, 1967; see this volume, section IV). For Wigner, this conception of symmetry principles is essentially related to our ignorance (if we could directly know all the laws of nature, we would not need to use symmetry principles in our search for them). Others, on the contrary, have arrived at a view according to which symmetry principles function as "transcendental principles" in the Kantian sense (see for instance Mainzer, 1996). It should be noted in this regard that Wigner's starting point, as quoted above, does not imply exact symmetries – all that is needed epistemologically is that the global symmetries hold approximately, for suitable spatiotemporal regions, such that there is sufficient stability and regularity in the events for the laws of nature to be discovered.

There is another reason why symmetries might be seen as being primarily epistemological. As we have mentioned, there is a close connection between the notions of symmetry and equivalence, and this leads also to a notion of irrelevance: the equivalence of space points (translational symmetry) is, for example, understood in the sense of the irrelevance of an absolute position to the physical description; in the case of local symmetries the irrelevant elements correspond to the presence of "surplus structure" in the theory.[19] There are two ways that one might interpret the epistemological significance of this: on the one hand, we might say that symmetries are associated with unavoidable redundancy in our *descriptions* of the world, while on the other hand we might maintain that symmetries indicate a limitation of our epistemic access – there are certain properties of objects, such as their absolute positions, that are not observable.

Finally, we would like to mention an aspect of symmetry that might very naturally be used to support either an ontological or an epistemological account. It is widely agreed that there is a close connection between symmetry and objectivity, the starting point once again being provided by spacetime symmetries: the laws by means of which we describe the evolution of physical systems have an objective validity because they are the same for all observers. The old and natural idea that what is objective should not depend upon the particular perspective under which it is taken into consideration is thus reformulated in the following group-theoretical terms: what is objective is what is invariant with respect to the transformation group of reference frames, or, quoting Hermann Weyl (1952, p. 132), "objectivity means invariance with respect to the group of automorphisms [of space-time]". The link between symmetry and objectivity is one theme of the paper by Kosso in section IV of this volume.

Summing up, symmetries in physics offer many interpretational possibilities, including ontological, epistemological and methodological. The position that one takes will depend in part on one's preferred approach to other issues in philosophy of science, including realism, the laws of nature, the relationship between mathematics and physics, the nature of theoretical entities, and so forth. It will also depend on whether one views symmetries as ultimately fundamental or derivative (be that in a methodological sense or, at the other extreme, an ontological sense). How to understand the status and significance of physical symmetries clearly presents a challenge to both physicists and philosophers.

---

[19] See Belot and Castellani, both this volume, section IV, and also Redhead's "surplus structure", this volume, section I.

# 4  *Structure of the book*

Our aim in this book is to provide a structured picture of the current philosophy of physics debate on symmetry, along with a context and framework for future debate and research in this field. As such, the aim is modest: there is no intention or aspiration to provide a comprehensive discussion of all philosophical issues that might arise from the roles of symmetries in physics. Rather, the content of this book clearly displays the issues that dominate the current discussion in philosophy of physics.

We have divided the book into four sections, each of which begins with a selection of classic texts from such authors as Leibniz, Kant, Curie, Weyl and Wigner. The first three sections of the book concern specific topics falling under the general heading of symmetry in physics: continuous symmetries, discrete symmetries, and symmetry breaking. Each contains a paper that reviews the current situation in the literature and highlights the main issues and controversies. The fourth section is devoted to the general interpretational questions arising in connection with symmetries.

*Section I:* Among the issues raised by Martin in his review of continuous symmetries, the one that dominates the papers that follow is a set of interrelated questions surrounding the interpretation of local symmetries. Martin himself spends some time addressing the status of the so-called "gauge principle", whose origins in Weyl's 1918 work – and particularly the philosophical background to this work – are the subject of Ryckman's paper. Brading and Brown pick up the historical thread with a discussion of Noether's 1918 theorems and the connection between symmetries and conservation laws. While their paper contains Noether's famous first theorem, concerning global symmetries, they also discuss the more complex case of local symmetries and the question of where the empirical significance of such symmetries lies. A theme common to the papers by Norton, Redhead, Earman, and Wallace is the "underdetermination problem" associated with theories containing local symmetries. These papers discuss how this problem arises (with respect both to the diffeomorphism freedom of GTR and to gauge theories), what interpretational problems follow, and how these may be tackled. The underdetermination problem is connected to the issue of which quantities in a local gauge theory should be interpreted as real. This problem is made particularly vivid by the Aharanov-Bohm effect; Nounou offers a discussion of this effect, in which she sets out her preferred approach based on the fibre bundle formulation of gauge theories – her paper contains a conceptual introduction to fibre bundles, designed to make the philosophical account accessible.

*Section II:* Under the general heading of discrete symmetries we find two distinct areas of research, each of which has a large associated literature. The first is permutation symmetry, reviewed by French and Rickles. The second is CPT, or rather, in fact, primarily P. In the philosophy of physics literature, parity (and parity violation) at the level of the fundamental laws has been the focus of attention, the absolute versus relational debate in the philosophy of space and time being the context. This is the topic of Pooley's review paper. Themes arising in these review papers are picked up by Huggett and Saunders. Huggett's first paper extends the French and Rickles discussion from bosons and fermions to other kinds of quantum particles ("quarticles"), while his second paper is a direct response to the discussion of handedness in Pooley's paper. Saunders advocates a version of Leibniz's Principle of the Identity of Indiscernibles that appeals to "weak discernibility", which is a natural generalization of Leibniz's law and, he argues, consistent with, and a useful tool with respect to, modern physics.

*Section III:* Philosophers have come to the topic of symmetry breaking only recently, and hence the main purpose of the review paper by Castellani is to provide an introduction and a framework for further work. In his "Rough Guide" Earman offers an approach to understanding symmetry breaking that makes use of the algebraic formulation of quantum theory, while Morrison's paper raises interpretational questions over the status of spontaneously broken symmetries.

*Section IV:* The final section contains a selection of papers by Ismael and van Fraassen, Belot, Kosso, and Castellani on general issues of the interpretation of symmetry. They pick up on issues ranging right across the material of the preceding sections, such as those of redundancy and surplus structure, symmetries of laws versus symmetries of solutions, and the relationship between symmetry and objectivity.

Our hope is that this volume will appeal to a wide audience, including philosophers of physics, philosophers of science, and physicists. It offers something for everyone who is curious about symmetries in physics, providing a research tool as well as a point of access into this fascinating area.